\documentclass[aps,twocolumn,showpacs]{revtex4}

\usepackage{graphicx}
\usepackage{epsfig}
\usepackage{amssymb}
\usepackage{amsmath}
\usepackage{psfrag}
\usepackage{color}

\newcommand{\br}{{\bf r}}

\newcommand{\sG}{{\mathrm{G}}}
\newcommand{\sS}{{\mathrm{S}}}
\newcommand{\sE}{{\mathrm{E}}}

\newcommand{\stp}{{\mathrm{p}}}	
\newcommand{\se}{{\mathrm{e}}}

\newcommand{\Imag}{\mathrm{Im} \, }
\newcommand{\Real}{\mathrm{Re} \, }

\newcommand{\hz}{\,\mathrm{s}^{-1}}

\newcommand{\mic}{\,\mathrm{\mu m}}

\begin{document}

\title{Strong Coupling to Two-Dimensional Anderson Localized Modes}
\author{A. Caz\'e, R. Pierrat and R. Carminati}
\email{remi.carminati@espci.fr}
\affiliation{Institut Langevin, ESPCI ParisTech, CNRS, 1 rue Jussieu,
75238 Paris Cedex 05, France}

\begin{abstract}
We use a scattering formalism to derive a condition of strong coupling between a resonant scatterer and an Anderson localized mode
for electromagnetic waves in two dimensions. The strong coupling regime is demonstrated based on exact numerical simulations, in perfect 
agreement with theory. The strong coupling threshold can be expressed in terms of the Thouless conductance and 
the Purcell factor. This connects key concepts in transport theory and cavity quantum electrodynamics, and provides a practical tool for the design 
or analysis of experiments.
\end{abstract}

\pacs{42.25.Dd,72.15.Rn,42.50.Pq}

\maketitle


Enhancing and controlling light-matter interaction has been an issue of tremendous interest for years.
The pioneering prediction of the dependance of the spontaneous decay rate of an emitter on its environment
in the weak coupling regime, known as the Purcell effect ~\cite{Purcell1946}, was observed in optics
by Drexhage~\cite{Drexhage1968}.  The development of cavity quantum electrodynamics (QED)
has led to the observation of the strong coupling regime, characterized by Rabi oscillations of the excited-state population, 
or a splitting in the frequency spectrum~\cite{CavityQED-books}. Strong coupling has been demonstrated with single atoms in 
engineered vacuum cavities~\cite{Rempe-Brune},  and in condensed matter using quantum-well or quantum-dot excitons in microcavities or 
photonic crystals~\cite{Weisbuch1992,Yoshie2004,Peter2005}.
In nanophotonics, surface-plasmon modes on metallic nanoparticles or substrates provide subwavelength light confinement
without a physical cavity, and strong coupling has been reported with quantum dots or molecules~\cite{Trugler2008,VanVlack2012,Chang2007,Bellessa2012,Lienau2013}. 
Multiple scattering in disordered media provides an alternative route since confined modes can be produced by the mechanism of Anderson localization~\cite{John1984-1987}.
Substantial modifications of the spontaneous decay rate (Purcell effect) have been demonstrated using quantum dots and localized modes in disordered photonics crystal 
waveguides~\cite{Lodahl2010}. In these one-dimensional structures, even fabrication imperfections in otherwise perfect waveguides
generate efficient localization on the micrometer scale~\cite{Lalanne2009-2012}, and the strong coupling regime is expected to be within 
experimental reach~\cite{Lodahl2012}. In addition to multiple scattering, near-field interactions also contribute to an enhancement of light-matter interaction
with large Purcell factors in the weak-coupling regime~\cite{Carminati2010-2011}.

In this Letter, we study the interaction between a resonant dipole scatterer and a two-dimensional (2D) Anderson localized mode,
based on a scattering formalism for electromagnetic waves. Using exact numerical simulations, we demonstrate the strong coupling regime in 2D localized systems.
The results are in perfect agreement with a simple coupled-mode theory. Using this theory, we examine the strong coupling criterion, and show that it can be 
expressed in terms of the Thouless conductance and the Purcell factor. This result establishes an interesting connection between concepts in 
transport theory and cavity QED. It also provides a simple rule for the design and/or the analysis of future experiments aiming at demonstrating or using 
(classical or quantum) strong coupling with electromagnetic waves.


In a first part, we use the LDOS spectrum to characterize an Anderson localized mode.
We consider a two-dimensional disordered medium and Transverse Electric (TE) waves (electric field perpendicular to the plane containing the 2D scatterers), so that
we are left with a scalar problem.
To introduce the methodology, let us first consider the canonical situation of a non-absorbing environment placed in a closed cavity. 
In this case, one can define an orthonormal discrete basis of eigenmodes with eigenfrequencies $\omega_n$ and eigenvectors $\se_n(\br)$.
The electromagnetic response of the medium can be expanded over the set of eigenmodes~\cite{MorseBook} 
\begin{equation}
\label{green_noabs}
\sG(\br,\br',\omega) = \sum_n c^2 \frac{\se_n^*(\br')\se_n(\br)}{\omega_n^2-\omega^2} \, ,
\end{equation}
where $c$ is the speed of light in vacuum, $\omega$ the frequency and $\sG(\br,\br',\omega)$ the outgoing 2D scalar Green function. 
In the general case of a leaky system the weak losses out of each mode can be taken into account phenomenologically using an effective linewidth $\Gamma_n$.
For an open system, as the one considered in this study, this effective linewidth accounts for radiative leakage. It could also account for other loss mechanisms, as
material absorption or out-of-plane scattering in a quasi 2D system. 
The electric part of the LDOS, relevant for the coupling with electric dipoles, is defined as $\rho(\br,\omega) = 2\omega/(\pi c^2)\,\Imag \sG(\br,\br,\omega)$~\cite{Joulain2003}.
Therefore the LDOS spectrum is given by
\begin{equation}
\label{ldos_localized}
\rho(\br,\omega) =\sum_n \rho_n(\br, \omega) = \sum_n \frac{A_n}{\pi} \frac{\Gamma_n/2}{(\omega_n-\omega)^2 + (\Gamma_n/2)^2} \, ,
\end{equation}
where $A_n = |\se_n(\br)|^2$.
The LDOS spectrum contains all the relevant parameters of a given mode (central frequency, linewidth and local intensity). A major interest is that it can in principle 
be determined experimentally from fluorescent lifetime measurements, even at the nanoscale in complex geometries~\cite{LDOS_SNOM}.
For convenience, we also define the Purcell factor associated to a given mode $n$, and a position $\br$, as $\mathrm{F}_p= \rho_n(\br, \omega_n)/\rho_0$, 
where $\rho_0=\omega/(2\pi c^2)$ is the vacuum LDOS in 2D.

\begin{figure}
\begin{center}
\includegraphics[width=6.5cm]{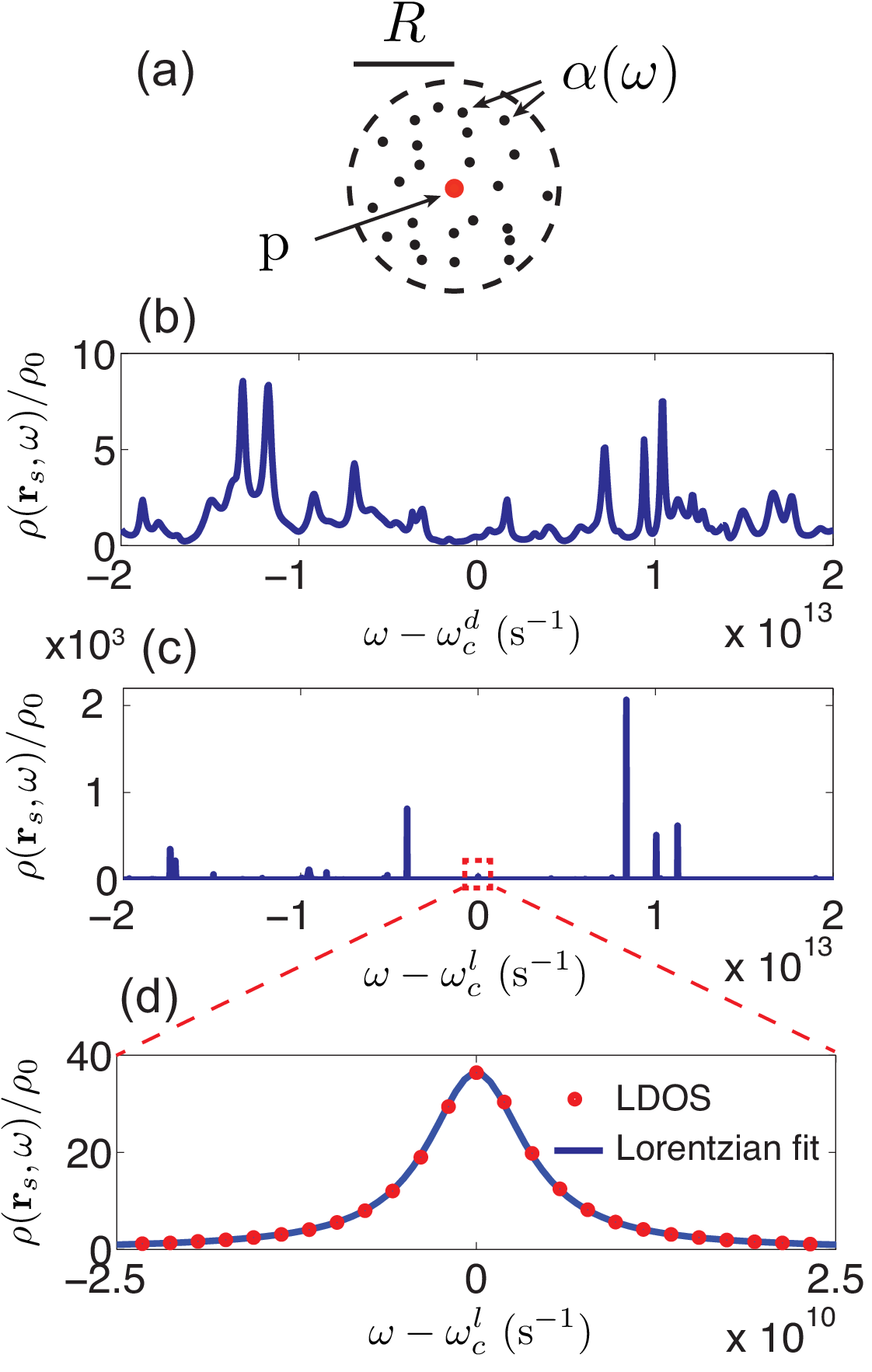}
\caption{\label{fig:anderson} (Color online) (a) Sketch of the system. (b) LDOS spectrum centered at $\omega_c^d = 2.7\times10^{15}\hz$ (diffusive regime). (c) LDOS spectrum centered at $\omega_c^l~=~1.5~\times~10^{15}\hz$ (localized regime). (d) Zoom on one peak in the localized regime. Circles correspond to a fit by Eq.~(\ref{ldos_localized}).}
\end{center}
\vspace{-0.5cm}
\end{figure}

In order to investigate Anderson localization numerically, we consider an assembly of 2D point scatterers described by their electric polarizability 
$\alpha(\omega) = (2\Gamma_0/k_0^2)(\omega_0-\omega-i\Gamma_0/2)^{-1}$, where $k_0=\omega/c$, $\omega_0$ is the resonance frequency and $\Gamma_0$ 
the natural linewidth. This form of the polarizability describes non absorbing scatterers and satisfies energy conservation. 
It has been chosen to provide the simplest model of a strongly scattering medium
exhibiting Anderson localization. The influence of absorption (in the host medium or in the scatterers) and of out-of-plane scattering (in a quasi 2D system)
 on Anderson localization is beyond the scope of the present study that is focused on the strong-coupling condition.
We have fixed $\omega_0 = 3\times 10^{15}\hz$ (visible optical radiation) and $\Gamma_0=5\times 10^{16}\hz \gg \omega_0$. 
With such a wide resonance, the scattering cross-section of the scatterers is constant over the spectral range considered in the numerical simulations below.
The scatterers are randomly distributed in a cylinder of radius $R$. In order to compute the LDOS at the central point $\br_S$, we need to compute the field
scattered at $\br_S$ when the system is illuminated by a source dipole $\stp$ also located at $\br_S$ [see Fig.~\ref{fig:anderson}(a)].
The exciting field on scatterer number $i$ is given by the self-consistent equation
\begin{equation}
\label{coupled_dipoles}
\sE_i = \mu_0\omega^2\sG_0(\br,\br_S,\omega)\stp + \frac{\omega^2}{c^2}\alpha(\omega)\sum_{j\ne i} \sG_0(\br_i,\br_j,\omega)\sE_j,
\end{equation}
where $\br_i$ is the position of scatterer number $i$.
The 2D vacuum Green function is $\sG_0(\br,\br',\omega) = (i/4) \operatorname{H}_0^{(1)}(k_0|\br-\br'|)$, where
$\operatorname{H}_0^{(1)}$ is the zero-order Hankel function of the first kind. 
For a system with $N$ scatterers, the linear system of $N$ self-consistent equations can be solved numerically. Once the exciting field on each scatterer is known, it is possible to compute the scattered field at $\br_S$, and to deduce the Green function and the LDOS $\rho(\br_S,\omega)$.

Let us consider one configuration of the random system, with $N=5000$ scatterers in a cylinder of radius $R=20\mic$.
Two computed LDOS spectra, with the same bandwidth but centered on two different central frequencies $\omega_c^d = 2.7\times10^{15}\hz$ (diffusive regime) and
 $\omega_c^l = 1.5\times10^{15}\hz$ (localized regime), are shown in Fig.~\ref{fig:anderson}(b) and \ref{fig:anderson}(c), respectively. 
To choose these two frequencies, we have estimated the localization length by $\xi = \ell_s\exp[\pi\Real(k_{\textrm{eff}}) \ell_s/2]$,
with $\ell_s$ the scattering mean free path and $k_{\textrm{eff}}$ the effective wavenumber in the medium~\cite{Gupta2003,Laurent2007}.
For a rough estimate, we have made the approximation $k_{\textrm{eff}}\approx k_0 + i/(2\ell_s)$, 
valid in the weak scattering limit.
In the spectrum shown in Fig.~\ref{fig:anderson}(b), one has $\xi \simeq 84 R$ and the sample is in the diffusive regime.
We observe a smooth profile corresponding to the intuitive picture of a continuum of modes.
Conversely, in Fig.~\ref{fig:anderson}(c), the localization length is $\xi \simeq R/5$ and the sample is in the localized regime.
We observe very sharp and well-separated peaks, each of them being a signature of a localized mode. A peaked spectrum,
characteristic of localized modes, is found numerically on any configuration of the disorder, provided that $\xi \ll R$.
A zoom on one of the LDOS peaks, as displayed in Fig.~\ref{fig:anderson}(d), shows that it can be perfectly fitted by a Lorentzian lineshape as in Eq.~(\ref{ldos_localized}),
demonstrating the relevance of this description. Such a Lorentzian lineshape for localized modes is consistent with measurements performed in disordered 
waveguides~\cite{Lodahl2010,Genack2011}. This isolated Anderson localized mode will be denoted by mode $M$ in the following, and will be used to demonstrate
numerically the strong coupling regime. It is characterized by an
eigenfrequency $\omega_M \simeq 1.5\times10^{15}\hz$, an effective linewidth $\Gamma_M \simeq 8\times10^9\hz$ (the quality factor $Q_M \simeq 1.8\times10^5$)
and a Purcell factor $\mathrm{F}_p \simeq 36$. 


In the second part, we describe theoretically the coupling between a resonant dipole scatterer and an Anderson localized mode, in order to establish the threshold condition for
strong coupling. In very general terms, the scattering medium is described by the scattered Green function $\sS(\br,\br',\omega) = \sG(\br,\br',\omega) - \sG_0(\br,\br',\omega)$, where 
$\sG_0$ is the vacuum Green function (or more generally the Green function in a reference medium). 
The resonant dipole scatterer, placed at position $\br_S$, is described by its electric polarizability $\alpha_S$ in vacuum (or in the same reference medium), such that its induced dipole
moment reads
\begin{equation}
\label{p_evol}
\stp(\omega) = \epsilon_0 \alpha_S(\omega) \sE^{\mathrm{exc}}(\br_S,\omega),
\end{equation}
where $\sE^{\mathrm{exc}}(\br_S,\omega)$ is the exciting field. The eigenmodes of the coupled systems are found by assuming that 
the exciting field is provided by the polarizable scatterer itself (no external illumination), so that
\begin{equation}
\label{e_evol}
\sE^{\mathrm{exc}}(\br_S,\omega) = \mu_0 \omega^2 \sS(\br_S,\br_S,\omega) \stp(\omega).
\end{equation}
Combining Eqs.~(\ref{p_evol}) and (\ref{e_evol}), one obtains the implicit equation satisfied by the eigenfrequencies of the coupled system~\cite{note}
\begin{equation}
\label{coupling_relation}
\frac{\omega^2}{c^2}\alpha_S(\omega)\sS(\br_S,\br_S,\omega) = 1.
\end{equation}
This general relation rules the coupling between the scatterer and its environment whatever the strength of this coupling (it is not restricted
to the strong-coupling regime).

In the case of an Anderson localized mode centered at $\omega_M$, the Green function in the vicinity of $\omega_M$ is given by 
\begin{equation}
\label{green_localized}
\sG(\br,\br',\omega) = \frac{c^2}{2\omega_M} \frac{\se_M^*(\br')\se_M(\br)}{\omega_M-\omega - i\Gamma_M/2} \, .
\end{equation}
The resonant scatterer, assumed on resonance with mode $M$, is described by a polarizability
\begin{equation}
\label{lorentzian_scatterer}
\alpha_S(\omega) = \frac{2c^2}{\omega^2}\frac{\Gamma^R_S}{\omega_M - \omega - i (\Gamma_S^R+\Gamma_S^{\textit{NR}})/2},
\end{equation}
where $\Gamma_S^R$ and $\Gamma_S^{\textit{NR}}$ are, respectively, the radiative and intrinsic non-radiative linewidth. This
polarizability describes either a classical resonant scatterer (the non-radiative linewidth corresponding to dissipation in the material),
or a quantum two-level system far from saturation (in this case $\Gamma_S^{\textit{NR}}=0$). Note that
$\Gamma_S^R$ also appears in the numerator as it includes the oscillator strength.
The complex eigenfrequencies $\Omega$ of the coupled system are solutions of Eq. (\ref{coupling_relation}), in which $\alpha_S(\omega)$ is given by 
Eq.~(\ref{lorentzian_scatterer}) and $\sS(\br_S,\br_S,\omega)$ is deduced from Eq.~(\ref{green_localized}). 
One finds two solutions
\begin{equation}
\label{eigenfrequencies}
\Omega^{\pm} = \omega_M \pm \left[g_c^2 - \frac{\{\Gamma_S^{\textit{NR}}-\Gamma_M\}^2}{16} \right]^{1/2} -\frac{i}{2}\left(\frac{\Gamma_M + \Gamma_S^{\textit{NR}}}{2}\right),
\end{equation}
where $g_c = (\Gamma_S^R\Gamma_M\mathrm{F}_p/4)^{1/2}$ is the coupling constant. Under the condition
\begin{equation}
\label{strong_coupling_criterion}
g_c^2 \ge \frac{(\Gamma_S^{\textit{NR}}-\Gamma_M)^2}{16},
\end{equation}
the two new eigenmodes of the coupled system are no longer degenerated. This defines the strong coupling regime.
For a quantum two-level system, $\Gamma_S^{\textit{NR}}=0$ and the condition is simply $g_c \ge \Gamma_M/4$, 
which is consistent with the usual criterion in cavity-QED~\cite{Yoshie2004,note_CQED}.
The frequency splitting between the two eigenmodes is given by the Rabi frequency $\Omega_R$, such that
\begin{equation}
\label{rabi_frequency}
\Omega_R = \frac{\Omega^+ - \Omega^-}{2} =  \left[g_c^2 - \frac{\{\Gamma_S^{\textit{NR}}-\Gamma_M\}^2}{16} \right]^{1/2}.
\end{equation}
The spectral width $\Gamma$ of the new eigenmodes is the average of $\Gamma_S^{\textit{NR}}$ and $\Gamma_M$, i.e., of the intrinsic linewidths of the uncoupled
system. Let us note that condition (\ref{strong_coupling_criterion}) is not sufficient to ensure that the Rabi splitting is larger than the linewidth (this would
be a necessary condition to observe Rabi oscillations in the time domain). One needs to satisfy the more restrictive condition $2\Omega_R\ge\Gamma$, that reads
\begin{equation}
\label{strong_coupling_observation_criterion}
g_c^2 \ge \frac{(\Gamma_S^{\textit{NR}})^2 + \Gamma_M^2}{8} \, .
\end{equation}
Finally let us note that the weak-coupling regime is recovered when Eq. (\ref{strong_coupling_criterion}) is not satisfied, in the limit $\Gamma_M\gg\Gamma_S$.
In this limit, Eq. (\ref{eigenfrequencies}) shows that the system remains degenerated. The resonance of the scatterer broadens without affecting the localized mode.
The broadening in this regime (or the change in the spontaneous decay rate for a quantum emitter) is exactly given by the Purcell factor.

\begin{figure}
\begin{center}
\includegraphics[width=8.5cm]{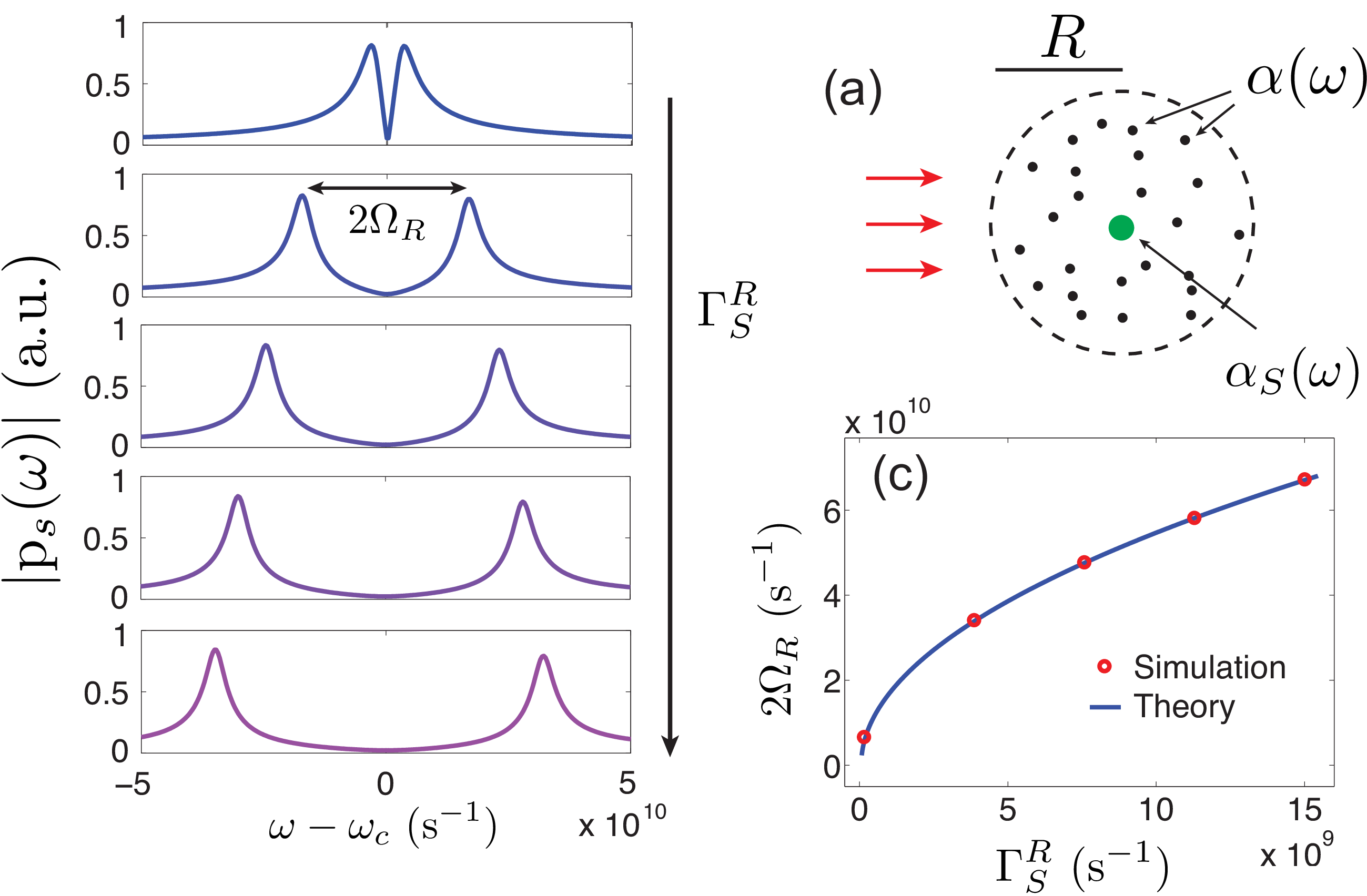}
\caption{(Color online) (a) Sketch of the system with a resonant probe scatterer placed at the center. (b) Spectra of the dipole moment $|\stp_S(\omega)|$ 
of the probe scatterer for different values of the radiative linewidth $\Gamma_S^R$ (from top to bottom $\Gamma_S^R = 1.5\times10^8\hz$; $3.9\times 10^9\hz$; $7.5\times 10^9\hz$; $11\times 10^9\hz$; $15\times10^9\hz$). (c) Frequency splitting in the spectrum of the dipole moment versus $\Gamma_S^R$. Solid line: Theoretical prediction by Eq.~(\ref{rabi_frequency}). Circles: Numerical simulations.}
\label{fig:splitting}
\end{center}
\vspace{-0.5cm}
\end{figure}

The expected strong coupling regime can be checked using exact numerical simulations. We consider the same system
as in Fig.~\ref{fig:anderson}(a), and add at position $\br_S$ a resonant dipole scatterer (probe scatterer), tuned to the resonance frequency $\omega_M$ of the
localized mode $M$ identified in the spectrum in Fig.~\ref{fig:anderson}(c).
The polarizability of the probe scatterer is given by Eq.~(\ref{lorentzian_scatterer}), with $\Gamma_S^{\textit{NR}} = 0$.
A sketch of the system is represented in Fig~\ref{fig:splitting}(a).
Under an illumination by an external field $\sE_0(\br,\omega)$ (plane wave),
a system of $N$ self-consistent equations similar to (\ref{coupled_dipoles}) can be written
\begin{eqnarray}
\label{coupled_dipoles_2}
\nonumber
\sE_i = \sE_0(\br_i,\omega) + \frac{\omega^2}{c^2}\alpha(\omega)\sum_{j\ne i} \sG_0(\br_i,\br_j,\omega)\sE_j \\*
 +  \frac{\omega^2}{c^2}\alpha_S(\omega) \sG_0(\br_i,\br_S,\omega)\sE_S,
\end{eqnarray}
where the exciting field $\sE_S$ on the probe scatterer is given by
\begin{equation}
\label{coupled_dipoles_3}
\sE_S = \sE_0(\br_S,\omega) + \frac{\omega^2}{c^2}\alpha(\omega)\sum_{j=1}^N \sG_0(\br_S,\br_j,\omega)\sE_j.
\end{equation}
Solving this linear systems with $N+1$ equations allows us to compute the induced dipole moment of the probe scatterer
$\stp_S(\omega) = \epsilon_0\alpha_S(\omega)\sE_S(\br_S,\omega)$.
We show in Fig.~\ref{fig:splitting}(b) the resulting spectrum for five different values of the radiative linewidth $\Gamma_S^R$ (increasing from top to bottom).
The Rabi splitting $2 \Omega_R$ increases with $\Gamma_S^R$, as expected from theory since the coupling strength $g_c$ scales as $(\Gamma_S^R)^{1/2}$.
The dependence of the Rabi splitting on $\Gamma_S^R$ extracted from the numerical simulations is shown in Fig.~\ref{fig:splitting}(c). Excellent
agreement is found with the theory. In summary, the simulations, performed without any approximation, have demonstrated the existence
of the strong coupling regime with an Anderson localized mode in two dimensions. The frequency splitting and its dependence on the parameters of the probe
scatterer are described quantitatively using the coupled-mode theoretical model, in which the parameters of the Anderson localized mode are extracted from a spectrum
of the LDOS.

We shall show that an alternative formulation of the strong coupling criterion can be given, that is particularly relevant in the case of
Anderson localization. Let us introduce the average linewidth of the electromagnetic modes $\delta\omega$ and the average mode spacing $\Delta\omega$.
Normalized linewidths $\hat{\Gamma}_S^R=\Gamma_S^R/\Delta\omega$ and $\hat{\Gamma}_M=\Gamma_M/\delta\omega$ can be introduced, for the probe scatterer 
and for the localized mode $M$. $\hat{\Gamma}_S^R=1$ means that the bandwidth of the scatterer covers on average only one mode of the disordered medium
(the linewidth of the resonant scatterer can be chosen or tuned to satisfy this condition).
When the probe scatterer is resonant with localized mode $M$, the strong coupling criterion given by Eq.~(\ref{strong_coupling_observation_criterion}) becomes
\begin{equation}
\mathrm{F}_p \geq \frac{1}{2}\, \frac{\hat{\Gamma}_M}{\hat{\Gamma}_S^R} \, g  \, ,
\label{inegalite_Fp_g}
\end{equation}
where $g= \delta\omega/\Delta\omega$ is the normalized Thouless conductance, a key concept in the theory of Anderson localization~\cite{Thouless}.
The localized regime corresponds to $g<1$ [this condition describes statistically a spectrum as that in Fig.~\ref{fig:anderson}(c)].
The inequality shows that the smaller the conductance, the smaller the critical Purcell factor permitting to enter the strong coupling regime. 
This confirms the idea that deeply localized modes in 2D or quasi-1D~\cite{Lalanne2009-2012,Lodahl2012} are particularly suitable to achieve 
strong coupling in the optical regime in condensed matter. 
For $\hat{\Gamma}_S^R \simeq 1$ and $\hat{\Gamma}_M \simeq 1$ (this condition is satisfied on average for the localized modes), the strong coupling criterion 
takes the remarkable simple form $\mathrm{F}_p \geq g/2$. This simple relation directly connects the Purcell factor (a central quantity in cavity QED) and the
Thouless conductance (a statistical concept in transport theory). Let us remark that the inverse of the Thouless conductance is statistically
the analogue of the finesse of a standard Fabry-P\'erot cavity that enters standard cavity QED analyses~\cite{CavityQED-books}.

In conclusion, we have demonstrated numerically the strong coupling regime between a resonant scatterer and an Anderson localized mode
for electromagnetic waves in two dimensions. The numerical results are in perfect agreement with a coupled-mode theory in which the parameters of 
the localized mode are extracted from a spectrum of the LDOS. The strong coupling threshold has been expressed in terms of the Thouless conductance and 
the Purcell factor. From the fundamental point of view, Eq.~(\ref{inegalite_Fp_g}) establishes an interesting connection between concepts in 
transport theory and cavity QED. On the practical side, it shows that once localization is reached ($g < 1$), the strong coupling criterion is not restrictive. 
For a resonant scatterer with a linewidth on the order of the averaged mode spacing (that on average is in coincidence with only one mode), 
the criterion is equivalent to having a Purcell factor $F_p > 1$. Although this criterion is rigorous only statistically, it provides a simple rule that could 
be useful in practice for the design and/or the analysis of future experiments aiming at demonstrating or using 
(classical or quantum) strong coupling with Anderson localized electromagnetic waves.

We acknowledge S. Bidault, Y. De Wilde, V. Krachmalnicoff,  J.J. S\'aenz, P. Sebbah, C. Vanneste and K.~Vynck for fruitful discussions. This work is supported by LABEX WIFI  (Laboratory of Excellence within the French Program "Investments for the Future") under references ANR-10-LABX-24 and ANR-10-IDEX-0001-02 PSL*.


\begin{thebibliography}{40}
\bibitem{note_CQED}
In the usual formulation, the explicit use of the transition dipole of the two-level system makes the criterion slightly different (but equivalent).
In our formalism the transition dipole is implicitly in the coupling constant $g_c$ through the radiative linewidth $\Gamma_S^R$.
\end{thebibliography}
\end{document}